\documentclass[twocolumn]{aastex701}

\begin{document}

\title{Constraining Gamma-Ray Emission Beyond the TOV Limit: A Targeted Fermi-GBM Search for Counterparts to 39 LVK Gravitational-Wave Candidates in GWTC-5.0}

\author[orcid=0009-0002-9856-6370,sname='Raj']{Navya Raj}
\altaffiliation{National Institute of Technology}
\affiliation{Department of Physics, National Institute of Technology Raipur, G.E. Road, Raipur, Chhattisgarh 492010, India}
\email[show]{navyaraj7903@gmail.com}
\begin{abstract}

We present a systematic search for sub-threshold gamma-ray counterparts within a one-day time window of 161 gravitational-wave (GW) events from the GWTC-5.0 catalog, whose inferred source masses exceed the adopted Tolman–Oppenheimer–Volkoff (TOV) limit, using the Gamma-ray Targeted Search (GTS) pipeline. The search identifies 39 candidate GW–gamma-ray pairs satisfying the adopted joint likelihood-ratio thresholds. None of these candidates exhibits a temporal offset compatible with the prompt-emission delay observed for GW170817 (\(\sim1.7\) s); the smallest temporal separation is 17 min 17.976 s, ruling out GW170817-like prompt temporal associations in our candidate sample. Several candidates nevertheless show substantial spatial consistency. In particular, the most spatially consistent pair has a joint localization overlap of 95.4\%, despite a temporal separation of 13 h 19 min 8.85 s, illustrating that high spatial agreement can occur by chance in the absence of temporal consistency. Among the sub-threshold gamma-ray candidates, we further identify an anomalous event with an exceptionally large single-instrument significance of \(\log\mathcal{L}\mathcal{R}=5250.84\). For comparison, the adopted \(3\sigma\) threshold corresponds to \(\log\mathcal{L}\mathcal{R}\approx9\). The extreme likelihood ratio of this candidate motivates further investigation of trigger sensitivity, onboard veto configurations, and the statistical interpretation of sub-threshold events. Overall, we find no physically viable multi-messenger associations between the GWTC-5.0 events and the searched gamma-ray candidates, and derive corresponding upper limits on prompt high-energy emission from GWTC-5.0 progenitors.

\end{abstract}



\section{Introduction}

The discovery of the binary neutron star merger GW170817 and the short gamma-ray burst GRB 170817A confirmed the connection between these two classes of events \citep{Abbott_2017_GW, Abbott_2017_MM}. GRB 170817A was very faint. If this event had occurred at a larger distance, standard onboard triggering algorithms on instruments like the Fermi Gamma-ray Burst Monitor (Fermi-GBM) would have missed it \citep{Abbott_2017_MM, Kocevski_2018}. Ground-based searches of archived continuous data are used to recover these weak, untriggered events \citep{Kocevski_2018, Yang_2020}.

These sub-threshold searches can significantly increase the short gamma-ray burst detection rate \citep{Cai_2025}. Correlating quiet gravitational-wave triggers with untriggered gamma-ray data reduces the background noise rate and increases the joint search volume \citep{Nitz_2019}. Likelihood-based searches of Swift-BAT continuous data have also recovered faint, off-axis bursts \citep{DeLaunay_2021}.

Most previous joint sub-threshold searches are limited to older gravitational-wave catalogs like GWTC-1 or GWTC-3 \citep{Nitz_2019, Hamburg_2020}. The newly released GWTC-5.0 catalog contains many compact binary merger candidates \citep{LVK_2026_GWTC5}. A systematic search for sub-threshold gamma-ray counterparts across this complete catalog using open-source pipelines has not been performed.

Cai et al. (2025) recently applied the ETJASMIN pipeline to perform joint, multi-instrument verifications of the Fermi-GBM sub-threshold catalog using GECAM observations \citep{Cai_2025}. In this study, we use the open-source Gamma-ray Targeted Search (GTS) pipeline \citep{GTS_2026} to perform a targeted search for sub-threshold counterparts around the trigger times and sky locations of the GWTC-5.0 candidates \citep{LVK_2026_GWTC5}. We restrict our analysis to Fermi-GBM continuous data occurring within a one-day window of these cataloged compact binary mergers \citep{LVK_2026_GWTC5}. This targeted follow-up strategy builds on coherent, multi-detector likelihood-search methods developed by Blackburn et al. (2015) and Kocevski et al. (2018) to search for weak gamma-ray signals in temporal and spatial coincidence with GW triggers (Goldstein et al. 2019; Hamburg et al. 2020). To maximize sensitivity, we apply the latest Fermi-GBM data-cleaning procedures, including removal of the 4–12 keV channels to reduce detector noise and Galactic transients. We analyze a 24-hour coincidence window around each GW trigger and identify 39 GW–GRB candidate pairs with non-trivial joint log-likelihood ratios. However, none satisfies the combined temporal, spatial, and astrophysical criteria for a credible multi-messenger association. Although the highest spatial-overlap candidate shows \(\sim94\%\) consistency between the GW and GRB localizations, its large temporal offset renders the association physically implausible. We therefore find no significant GW–GRB association and use the candidate population to characterize the sub-threshold background and constrain prompt high-energy emission from the GW sources.

\section{SAMPLE}
We select a sample of 161 gravitational-wave (GW) events from the GWTC-5.0 catalog\footnote{\url{https://gwosc.org/eventapi/html/allevents/}}, supplemented by one high-significance candidate (GW250114\_082203) from a preliminary O4 release. Because a strict prompt coincidence search yields too few events to robustly characterize spatiotemporal correlations, we expand the search window to 24 hours around each GW trigger to increase the candidate sample, yielding 58 candidate GW--GRB pairs. We exclude 19 pairs whose GW triggers lacked parameter-estimation sky maps due to high false alarm rates ($\text{FAR} > 1\,\mathrm{yr}^{-1}$ across all pipelines), leaving 39 candidate pairs for full GTS processing.

For each of these 39 events, we adopt the \texttt{IMRPhenomXPHM\_SpinTaylor} sky localization map and identify coincident sub-threshold gamma-ray bursts from Fermi-GBM\footnote{\url{https://gcn.gsfc.nasa.gov/gcn/fermi_gbm_subthresh_archive.html}} within the 24-hour window. For each burst, we retrieve continuous Time-Tagged Event (TTE) data and spacecraft position-history files corresponding to the reported trigger time, $T_{\text{GRB}}$ \citep{Meegan_2009}. We process these data using the GTS pipeline to compute coherent log-likelihood ratios quantifying candidate significance \citep{USRA_2026}, followed by a joint likelihood analysis evaluating temporal delays and spatial overlaps \citep{Nitz_2019, Hamburg_2020}.

\begin{table}
\centering
\caption{Sub-threshold GRB candidates within the 24-hour analysis window, one row per GW-GRB pair, ordered by decreasing log-likelihood ratio. GRB Type is the targeted-search duration-based classification ($T_{\rm cand} < 2\,$s = Short GRB), not T90-based.}
\label{tab:grb_info}
\begin{tabular}{r r r l r}
\hline
ID & Trigger & $\log\mathcal{L}_{\rm ratio}$ & GRB Type & Search Timescale (s) \\
\hline
1 & 740166710 & 5250.84 & Short GRB & 0.128 \\
2 & 740166710 & 5250.84 & Short GRB & 0.128 \\
3 & 759019271 & 284.0 & Short GRB & 0.256 \\
4 & 759019271 & 284.0 & Short GRB & 0.256 \\
5 & 754776177 & 258.03 & Short GRB & 0.064 \\
6 & 758917876 & 244.18 & Short GRB & 0.128 \\
7 & 758917876 & 244.18 & Short GRB & 0.128 \\
8 & 758917876 & 244.18 & Short GRB & 0.128 \\
9 & 747661355 & 119.34 & Short GRB & 0.256 \\
10 & 755576606 & 79.9 & Short GRB & 0.512 \\
11 & 754570926 & 78.87 & Short GRB & 0.064 \\
12 & 754570926 & 78.87 & Short GRB & 0.064 \\
13 & 747000169 & 77.65 & Short GRB & 0.256 \\
14 & 746731612 & 63.94 & Short GRB & 0.128 \\
15 & 749409554 & 63.3 & Long GRB & 8.192 \\
16 & 749409554 & 63.3 & Long GRB & 8.192 \\
17 & 746684112 & 46.29 & Short GRB & 0.128 \\
18 & 746197881 & 44.81 & Long GRB & 4.096 \\
19 & 748190210 & 41.46 & Short GRB & 0.064 \\
20 & 748985632 & 35.84 & Short GRB & 0.064 \\
21 & 757181763 & 22.66 & Long GRB & 8.192 \\
22 & 741848134 & 18.88 & Long GRB & 2.048 \\
23 & 754439562 & 18.4 & Short GRB & 0.128 \\
24 & 746739456 & 12.9 & Short GRB & 0.064 \\
25 & 748414684 & 12.03 & Short GRB & 0.512 \\
26 & 748828348 & 10.46 & Long GRB & 2.048 \\
27 & 758135659 & 10.39 & Short GRB & 0.064 \\
28 & 758135659 & 10.39 & Short GRB & 0.064 \\
29 & 754251071 & 9.01 & Long GRB & 2.048 \\
30 & 741908097 & 8.78 & Long GRB & 4.096 \\
31 & 742852188 & 7.96 & Short GRB & 0.064 \\
32 & 734754392 & 7.95 & Short GRB & 0.064 \\
33 & 746311162 & 7.49 & Short GRB & 0.064 \\
34 & 738941649 & 7.27 & Short GRB & 0.064 \\
35 & 738941649 & 7.27 & Short GRB & 0.064 \\
36 & 758771314 & 7.18 & Short GRB & 0.064 \\
37 & 748992037 & 6.63 & Short GRB & 0.128 \\
38 & 748640542 & 6.1 & Short GRB & 0.512 \\
39 & 741891537 & 5.81 & Long GRB & 4.096 \\
\hline

\end{tabular}

\end{table}
\section{Methodology}
\vspace{0.5em}
\subsection{Fermi-GBM Data Search Configuration}
\vspace{0.5em}
We searched the Fermi Gamma-ray Burst Monitor (Fermi-GBM) data using GTS pipeline. The analysis utilizes all 14 GBM detectors, comprising 12 sodium iodide (NaI) and two bismuth germanate (BGO) scintillation detectors \citep{Meegan_2009}. To reduce the contribution from electronic noise and poorly constrained extreme-energy channels, the native detector energy channels were rebinned using the predefined energy-channel boundaries adopted by the GTS analysis \citep{Kocevski_2018, Hamburg_2020}. The NaI and BGO detectors were analyzed over their respective GTS search energy ranges.

For each GW event, continuous Fermi-GBM data were searched within a 24-hour window to identify candidate sub-threshold gamma-ray bursts. To evaluate the statistical significance of each candidate, we performed a coherent likelihood analysis over a 10-second interval ($T_0 - 5\,\mathrm{s}$ to $T_0 + 5\,\mathrm{s}$) centered on the sub-threshold GRB trigger time, $T_0$ (listed under the {Trigger} column in Tables~\ref{tab:grb_info} and \ref{tab:gw_grb_assoc}).

Within this 10-second interval, the pipeline evaluated eight logarithmically spaced timescales ranging from $0.064\,\mathrm{s}$ to $8.192\,\mathrm{s}$ \citep{Kocevski_2018}. For each timescale, the search window was evaluated at eight phase offsets separated by one-eighth ($12.5\%$) of the corresponding bin width, resulting in an $87.5\%$ overlap between adjacent search bins. This phased sliding-window procedure prevents short-duration signals from being missed or split across bin boundaries \citep{Goldstein_2016, Kocevski_2018}. All combinations of time, duration, and phase were evaluated coherently across the Fermi-GBM detector array.

When the likelihood search yielded multiple candidates exceeding our baseline threshold of $\log\mathcal{L}\mathcal{R} \ge 5$ across different timescales or phase shifts within the 10-second window, we retained only the single candidate with the highest log-likelihood ratio ($\log\mathcal{L}\mathcal{R}$) for a given sub-threshold GRB trigger ($T_0$).
\subsubsection{Background Modeling and Likelihood Analysis}
\vspace{0.5em}
The GBM background varies with time due to contributions from trapped particles, Earth occultation, and cosmic-ray-induced events \citep{vonKienlin_2020, Antier_2019}. For each detector and energy channel, the background is estimated using data from intervals surrounding, but excluding, the search window. A smooth polynomial model is fitted to these background intervals to estimate the expected background count rate, $\mathcal{B}_i$, and its variance \citep{Goldstein_2016, Kocevski_2018}. This approach minimizes the possibility of a transient signal contaminating the background estimate.

For each sliding time window, coherent log-likelihood ratios are evaluated across all 14 GBM detectors by comparing the signal-plus-background hypothesis with the background-only null hypothesis \citep{Blackburn_2015, Cai_2025}. The signal model is represented by three spectral templates corresponding to soft, normal, and hard spectral shapes, which are folded through the energy- and direction-dependent detector response. The likelihood ratio is defined as
\begin{equation}
\mathrm{LLR} = \ln \left[ \frac{\mathcal{L}(D \mid S + B)}{\mathcal{L}(D \mid B)} \right],
\end{equation}
where $D$ represents the observed detector data, $S$ the transient signal, and $B$ the estimated background. For each trial interval, the spectral template and signal amplitude that maximize the LLR are selected.

To improve computational efficiency, candidate intervals are first pre-filtered using a preliminary threshold of $\log \mathrm{LR} \ge 5$ \citep{Goldstein_2016, Hamburg_2020}. Only intervals exceeding this threshold are subjected to the subsequent computationally intensive likelihood optimization. This threshold is used solely as a pre-filter and is not considered a final detection threshold.

Under the assumptions of Wilks' theorem, the likelihood-ratio statistic approximately follows a $\chi^2$ distribution with one degree of freedom. In this framework, $\log \mathrm{LR} \approx 9$ corresponds to a nominal $3\sigma$ significance level \citep{Kocevski_2018, Cai_2025}. We use the LLR values to rank the GRB candidates, while their temporal and spatial consistency with the corresponding GW events is assessed independently.

\subsection{GBM Localization and Systematic Uncertainty}
\vspace{0.5em}
For candidates passing the initial likelihood selection, the source direction is determined by evaluating the likelihood over a grid of trial sky positions using the direction- and energy-dependent response of the GBM detectors. The resulting likelihood surface is converted into a relative probability map and transformed from spacecraft coordinates to equatorial (J2000) coordinates using the spacecraft position and attitude information \citep{Blackburn_2015, Goldstein_2016, Kocevski_2018}.
The statistical localization is further corrected for systematic uncertainties arising from the detector response, spacecraft geometry, and photon scattering \citep{Connaughton_2016, Goldstein_2017}. The localization map is convolved with an empirically calibrated point-spread function described by a core-plus-tail model \citep{Connaughton_2016}.

\subsection{GW--GRB Spatial Association}
\vspace{0.5em}
For GW events with an available gravitational-wave sky map, $P_{\mathrm{GW}}(\hat{n})$, we compare it with the systematic-corrected GBM localization, $P_{\mathrm{GRB}}^{\mathrm{sys}}(\hat{n})$. Their spatial consistency is quantified using the overlap integral
\begin{equation}
\mathcal{A} = \int_{4\pi} P_{\mathrm{GRB}}^{\mathrm{sys}}(\hat{n}) P_{\mathrm{GW}}(\hat{n})\,d\Omega .
\end{equation}
A larger $\mathcal{A}$ indicates greater spatial consistency between the two maps \citep{Abbott_2017_MM, Pillas_2023}.
For candidates satisfying the adopted spatial-association criterion, the two maps are combined by pixel-wise multiplication and normalization:
\begin{equation}
P_{\mathrm{joint}}(\hat{n}) = \frac{P_{\mathrm{GRB}}^{\mathrm{sys}}(\hat{n}) P_{\mathrm{GW}}(\hat{n})}{\int_{4\pi} P_{\mathrm{GRB}}^{\mathrm{sys}}(\hat{n}') P_{\mathrm{GW}}(\hat{n}')\,d\Omega'},
\end{equation}
where $P_{\mathrm{joint}}(\hat{n})$ is the normalized joint probability at sky direction $\hat{n}$, $P_{\mathrm{GRB}}^{\mathrm{sys}}(\hat{n})$ is the GBM GRB probability map after accounting for systematic uncertainties, and $P_{\mathrm{GW}}(\hat{n})$ is the gravitational-wave sky-probability map. The denominator is the normalization factor, obtained by integrating the product of the two probability maps over the full sky, ensuring that the joint probability integrates to unity. Here, $\hat{n}$ denotes a sky direction and $d\Omega$ is the corresponding solid-angle element.
This gives a joint localization based on the two independent sky maps. The combination is performed at the map level and is not a joint fit to the GBM photon data and GW strain data \citep{Abbott_2017_MM, Ashton_2018, Pillas_2023}.
\begin{figure}[ht!]
    \centering

    \includegraphics[width=\columnwidth]{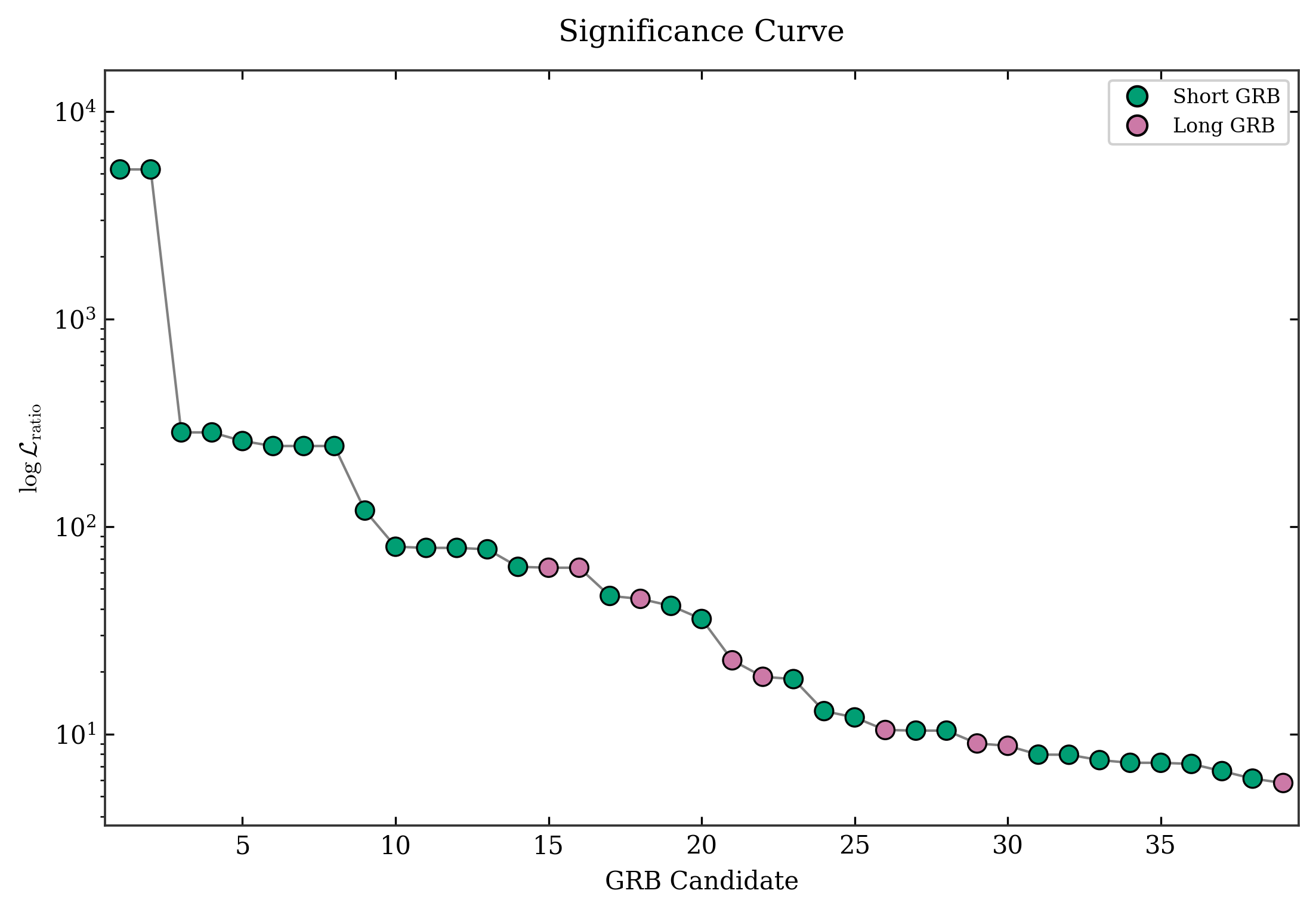}

    \vspace{0.2cm}

    \includegraphics[width=\columnwidth]{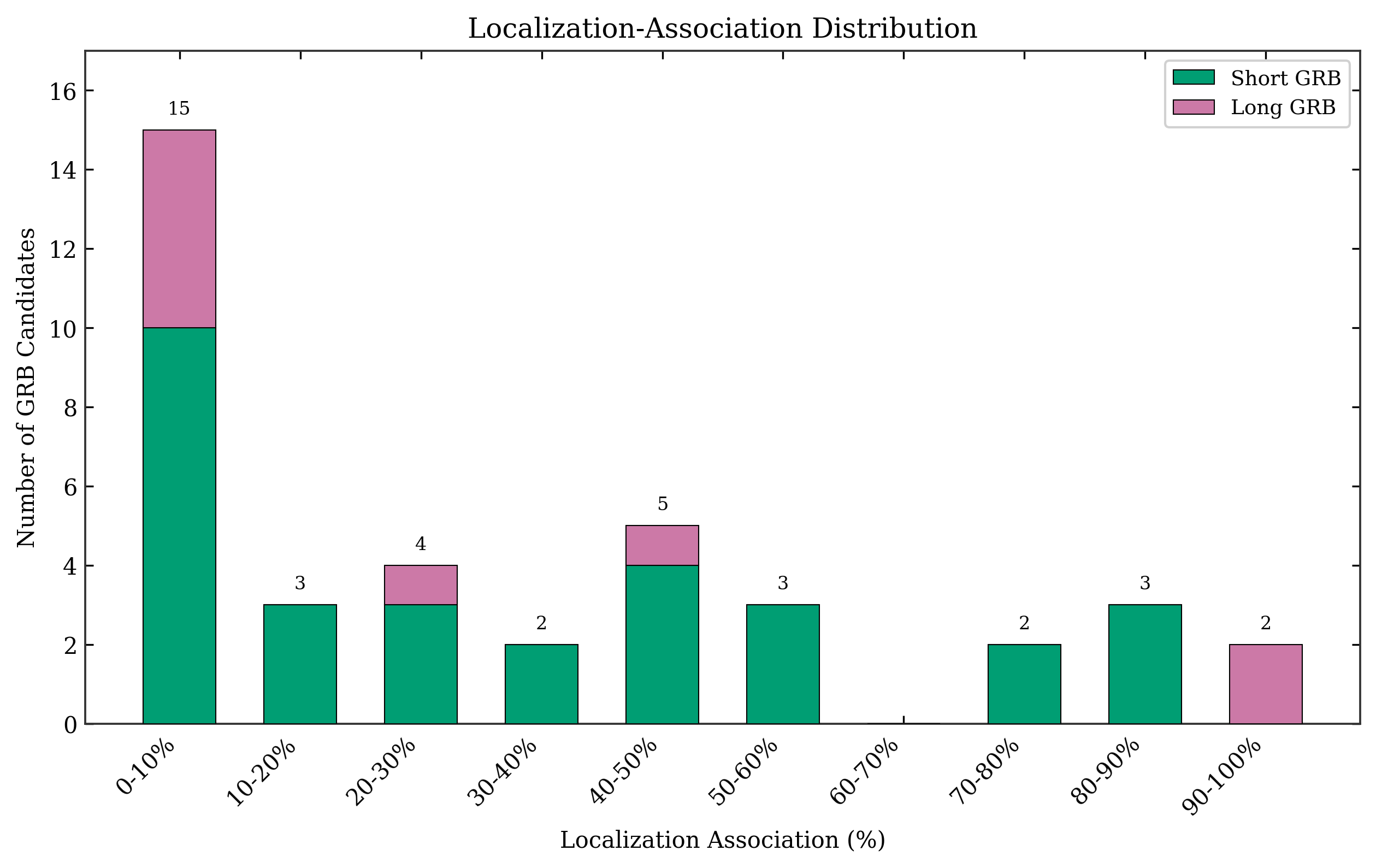}

    \caption{Results of the sub-threshold gamma-ray search. 
    Panel (a) shows the significance distribution of the GW--GRB candidates, 
    while panel (b) shows the distribution of the spatial association between 
    the GW and GRB localization regions.}
    \label{fig:results}
\end{figure}
\begin{figure}[!t]
    \centering
    \includegraphics[width=\columnwidth]{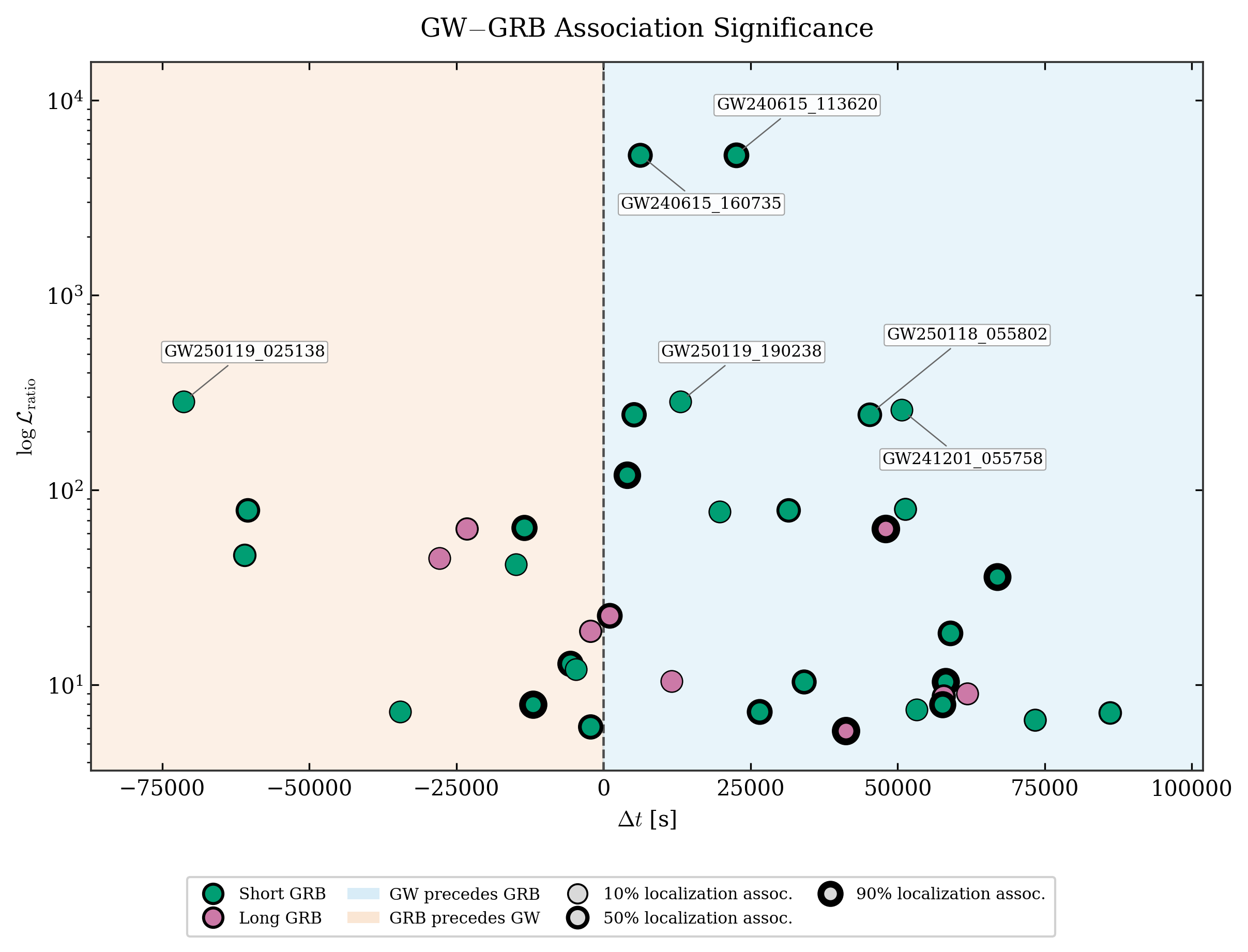}
    \caption{
        Spatial association between the gravitational-wave and
        gamma-ray burst localization regions for the candidate events.
    }
    \label{fig:association}
\end{figure}
\begin{figure*}[htbp]
    \centering
    \includegraphics[width=\linewidth]{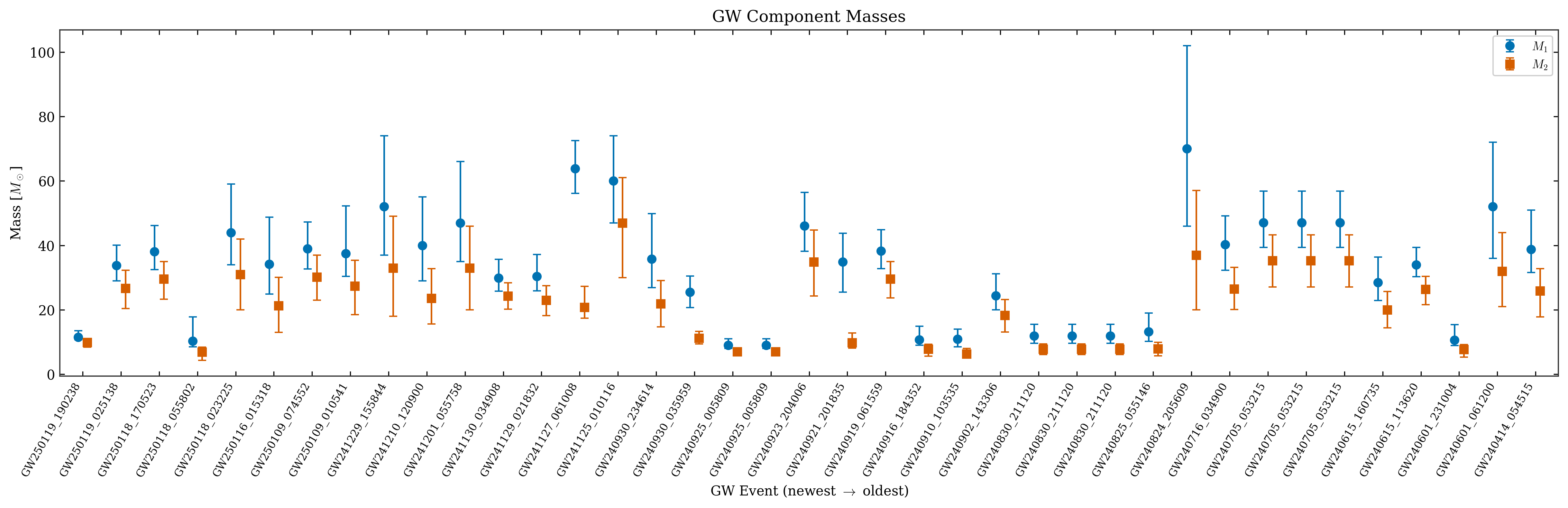}
    \caption{Inferred component masses of the compact-object binaries for the GW events in our sample. For each event, blue circles and orange squares denote the primary ($m_1$) and secondary ($m_2$) source-frame component masses, respectively, with error bars indicating the 90\% credible intervals from the corresponding GW parameter-estimation analysis. Events are ordered along the $x$-axis as listed in Table~\ref{tab:gw_grb_assoc}. With component masses ranging from $\sim 7\,M_\odot$ up to $\sim 100\,M_\odot$ (including uncertainties), all events in the sample lie well above the maximum non-rotating neutron star mass permitted by the TOV limit ($\sim 2$--$3\,M_\odot$), indicating that the sample is consistent with a population of binary black hole (BBH) mergers rather than binaries containing neutron stars.}
    \label{fig:figure3_component_masses}
\end{figure*}
\begin{figure}[!t]
    \centering
    \includegraphics[width=\columnwidth]{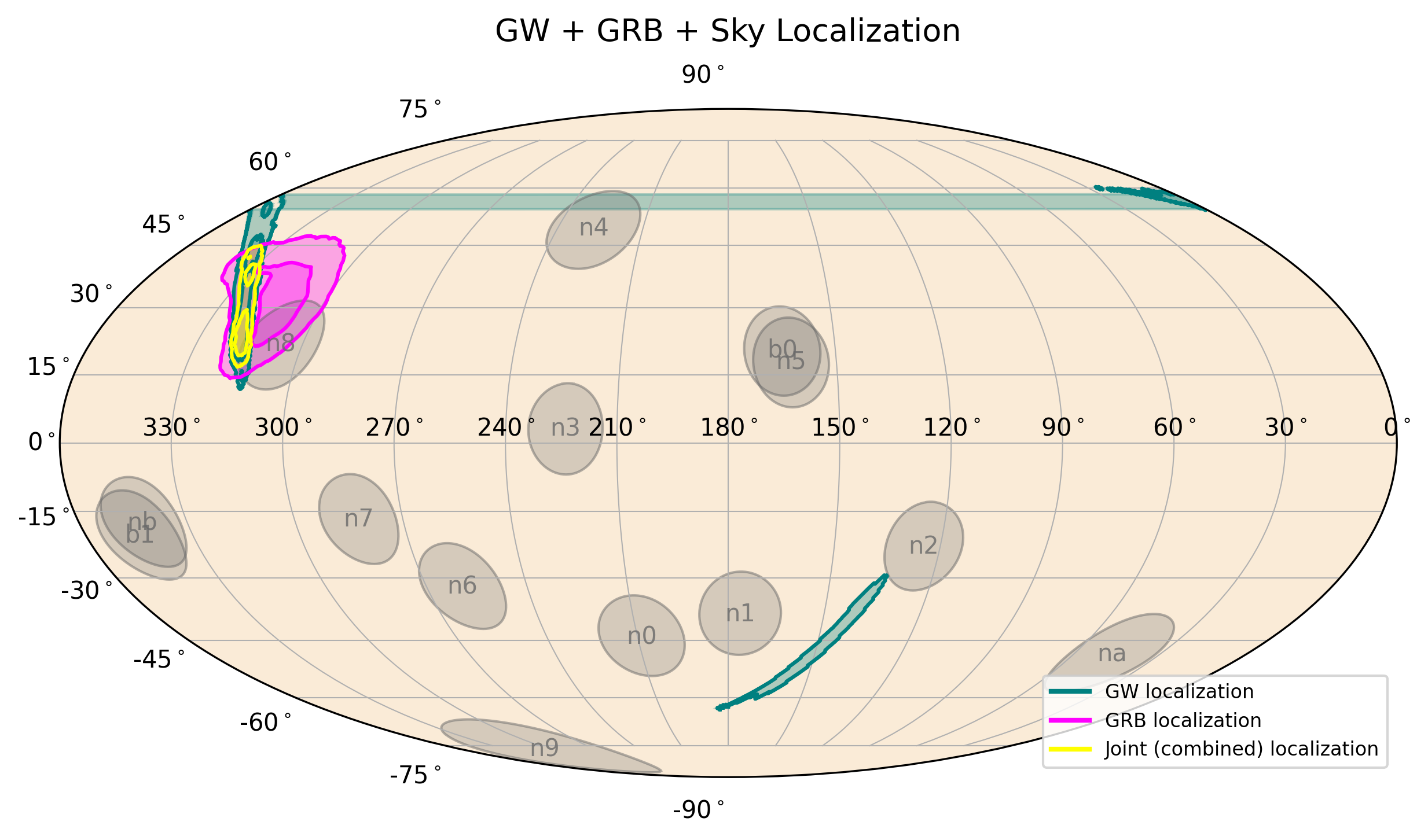}
    \caption{
    Sky localization overlap between the GW and GRB candidates on an
    equatorial Mollweide projection. Gray ellipses mark the GBM
    detectors' fields of view at the trigger time. For candidate
    GW240930\_035959 (Fermi-GBM trigger 749409554), the pipeline finds
    a 95.4\% spatial association, higher than the 93.7\% found for
    the confirmed GW170817/GRB~170817A pair using the same method.
    }
    \label{fig:highestassociation}
\end{figure}
\section{Results}
\label{sec:results}
\begin{table*}[t]
\centering
\caption{GW-GRB candidate associations within the 24-hour analysis window, sorted reverse-chronologically (newest first). $\Delta t$ is shown in its original human-readable format; Preceding Event indicates which of GW/GRB occurred first.}
\label{tab:gw_grb_assoc}
\begin{tabular}{r l r r l l l l}
\hline
ID & GW Name & Trigger & Localization Association (\%) & $|\Delta t|$ & Preceding Event & $M_1\,(M_\odot)$ & $M_2\,(M_\odot)$ \\
\hline
3 & GW250119\_190238 & 759019271 & 0.0 & 3h 38m 33.192s & GW & $11.54^{+1.95}_{-0.81}$ & $9.95^{+0.77}_{-1.56}$ \\
4 & GW250119\_025138 & 759019271 & 0.1 & 19h 49m 33.192s & GRB & $33.8^{+6.3}_{-4.8}$ & $26.7^{+5.6}_{-6.3}$ \\
7 & GW250118\_170523 & 758917876 & 41.7 & 1h 25m 53.896s & GW & $38.1^{+8.1}_{-5.6}$ & $29.6^{+5.4}_{-6.3}$ \\
6 & GW250118\_055802 & 758917876 & 29.8 & 12h 33m 14.896s & GW & $10.3^{+7.5}_{-1.8}$ & $6.9^{+1.5}_{-2.6}$ \\
8 & GW250118\_023225 & 758917876 & 16.3 & 12h 33m 14.896s & GW & $44^{+15}_{-10}$ & $31^{+11}_{-11}$ \\
36 & GW250116\_015318 & 758771314 & 12.3 & 23h 55m 16.384s & GW & $34.2^{+14.5}_{-9.3}$ & $21.3^{+8.8}_{-8.3}$ \\
27 & GW250109\_074552 & 758135659 & 36.6 & 9h 28m 27.704s & GW & $39.0^{+8.3}_{-6.3}$ & $30.2^{+6.8}_{-7.2}$ \\
28 & GW250109\_010541 & 758135659 & 88.1 & 16h 8m 38.704s & GW & $37.5^{+14.7}_{-7.1}$ & $27.4^{+8.0}_{-8.9}$ \\
21 & GW241229\_155844 & 757181763 & 49.2 & 17m 17.976s & GW & $52^{+22}_{-15}$ & $33^{+16}_{-15}$ \\
10 & GW241210\_120900 & 755576606 & 4.2 & 14h 14m 26.640s & GW & $40^{+15}_{-11}$ & $23.6^{+9.2}_{-8.0}$ \\
5 & GW241201\_055758 & 754776177 & 0.0 & 14h 4m 59.640s & GW & $47^{+19}_{-12}$ & $33^{+13}_{-13}$ \\
12 & GW241130\_034908 & 754570926 & 29.0 & 16h 47m 1.872s & GRB & $29.9^{+5.8}_{-4.1}$ & $24.3^{+4.1}_{-4.1}$ \\
11 & GW241129\_021832 & 754570926 & 27.3 & 8h 43m 34.128s & GW & $30.4^{+6.8}_{-4.5}$ & $23.0^{+4.5}_{-4.8}$ \\
23 & GW241127\_061008 & 754439562 & 47.3 & 16h 22m 34.768s & GW & $63.8^{+8.7}_{-7.7}$ & $20.8^{+6.5}_{-3.4}$ \\
29 & GW241125\_010116 & 754251071 & 3.4 & 17h 9m 53.976s & GW & $60^{+14}_{-13}$ & $47^{+14}_{-17}$ \\
16 & GW240930\_234614 & 749409554 & 7.4 & 6h 27m 6.144s & GRB & $35.8^{+14.0}_{-8.9}$ & $21.9^{+7.2}_{-7.2}$ \\
15 & GW240930\_035959 & 749409554 & 95.4 & 13h 19m 8.856s & GW & $25.5^{+5.0}_{-4.8}$ & $11.2^{+2.1}_{-1.8}$ \\
20 & GW240925\_005809 & 748985632 & 84.1 & 18h 35m 43.960s & GW & $9.02^{+1.97}_{-0.99}$ & $6.99^{+0.83}_{-1.27}$ \\
37 & GW240925\_005809 & 748992037 & 4.3 & 20h 22m 28.768s & GW & $9.0^{+2.0}_{-1.0}$ & $7.0^{+0.8}_{-1.3}$ \\
26 & GW240923\_204006 & 748828348 & 0.6 & 3h 12m 21.488s & GW & $46.1^{+10.3}_{-7.9}$ & $34.9^{+9.9}_{-10.6}$ \\
38 & GW240921\_201835 & 748640542 & 41.0 & 36m 12.488s & GRB & $34.9^{+8.9}_{-9.4}$ & $9.8^{+3.0}_{-1.7}$ \\
25 & GW240919\_061559 & 748414684 & 0.0 & 1h 17m 55.128s & GRB & $38.3^{+6.6}_{-5.5}$ & $29.6^{+5.4}_{-5.9}$ \\
19 & GW240916\_184352 & 748190210 & 0.0 & 4h 7m 1.936s & GRB & $10.7^{+4.2}_{-1.7}$ & $7.8^{+1.5}_{-2.2}$ \\
9 & GW240910\_103535 & 747661355 & 78.2 & 1h 7m 0.384s & GW & $10.9^{+3.1}_{-2.4}$ & $6.3^{+1.7}_{-1.3}$ \\
13 & GW240902\_143306 & 747000169 & 0.9 & 5h 29m 43.576s & GW & $24.4^{+6.8}_{-4.4}$ & $18.3^{+4.9}_{-5.2}$ \\
24 & GW240830\_211120 & 746739456 & 58.0 & 1h 33m 43.360s & GRB & $11.9^{+3.6}_{-2.3}$ & $7.8^{+1.6}_{-1.7}$ \\
14 & GW240830\_211120 & 746731612 & 56.0 & 3h 44m 27.232s & GRB & $11.9^{+3.6}_{-2.3}$ & $7.8^{+1.6}_{-1.7}$ \\
17 & GW240830\_211120 & 746684112 & 10.2 & 16h 56m 7.744s & GRB & $11.9^{+3.6}_{-2.3}$ & $7.8^{+1.6}_{-1.7}$ \\
33 & GW240825\_055146 & 746311162 & 0.0 & 14h 47m 36.832s & GW & $13.2^{+5.8}_{-3.0}$ & $7.9^{+2.0}_{-2.2}$ \\
18 & GW240824\_205609 & 746197881 & 0.0 & 7h 44m 49.536s & GRB & $70^{+32}_{-24}$ & $37^{+20}_{-17}$ \\
31 & GW240716\_034900 & 742852188 & 71.3 & 16h 0m 48.576s & GW & $40.3^{+8.8}_{-8.0}$ & $26.5^{+6.7}_{-6.4}$ \\
39 & GW240705\_053215 & 741891537 & 92.1 & 11h 26m 40.976s & GW & $47.1^{+9.7}_{-7.7}$ & $35.3^{+8.0}_{-8.2}$ \\
30 & GW240705\_053215 & 741908097 & 21.5 & 16h 2m 41.488s & GW & $47.1^{+9.7}_{-7.7}$ & $35.3^{+8.0}_{-8.2}$ \\
22 & GW240705\_053215 & 741848134 & 7.3 & 36m 41.512s & GRB & $47.1^{+9.7}_{-7.7}$ & $35.3^{+8.0}_{-8.2}$ \\
2 & GW240615\_160735 & 740166710 & 35.8 & 1h 44m 15.000s & GW & $28.5^{+7.9}_{-5.6}$ & $20.0^{+5.7}_{-5.6}$ \\
1 & GW240615\_113620 & 740166710 & 48.2 & 6h 15m 30.000s & GW & $34.0^{+5.4}_{-3.7}$ & $26.4^{+4.0}_{-4.8}$ \\
35 & GW240601\_231004 & 738941649 & 0.6 & 9h 35m 54.424s & GRB & $10.6^{+4.8}_{-1.7}$ & $7.7^{+1.5}_{-2.4}$ \\
34 & GW240601\_061200 & 738941649 & 50.8 & 7h 22m 9.576s & GW & $52^{+20}_{-16}$ & $32^{+12}_{-11}$ \\
32 & GW240414\_054515 & 734754392 & 89.9 & 3h 18m 43.384s & GRB & $38.8^{+12.1}_{-7.2}$ & $25.9^{+6.9}_{-8.1}$ \\
\hline
\end{tabular}
\end{table*}

Applying the GTS pipeline to the 161 GWTC-5.0 events
recovered 39 candidate GW--GRB pairs, none of which exhibit
temporal offsets consistent with a prompt physical counterpart.
In Figure~\ref{fig:results}, Panel (a) ranks all 39 candidates by log-likelihood ratio; two candidates reach $\log \mathcal{L} \approx 5\times10^{3}$, roughly an order of magnitude above the remaining sample, which declines smoothly from $\sim300$ to $\sim6$. Panel (b) bins candidates by localization association: 15 (10 short, 5 long) fall in the lowest bin (0--10\%), only 3 bins exceed 20 candidates combined above 70\%, and just 2 candidates exceed 90\% association.
Figure~\ref{fig:association} shows the log-likelihood ratio
($\log \mathcal{L}_{\mathrm{ratio}}$) plotted against the time separation,
$\Delta t = T_{\mathrm{GRB}} - T_{\mathrm{GW}}$, for all 39 GW--GRB
candidate pairs within the 24-hour search window. Here, $T_{\mathrm{GRB}}$
is derived from the Fermi-GBM trigger MET and the candidate start-time
offset, $T_{\mathrm{GRB}} = T_0 + t_{\mathrm{start}}$, $t_{\mathrm{start}}$ is the time offset, in seconds, within the search window at which the targeted search identifies the onset of the candidate GRB signal relative to $T_0$.
while $T_{\mathrm{GW}}$ is the GPS trigger time of the GW event, with
both times converted to a common UTC reference before differencing. The
vertical dashed line marks $\Delta t = 0$, separating the
GRB-precedes-GW region (orange, $\Delta t < 0$) from the GW-precedes-GRB
region (blue, $\Delta t > 0$). Marker color denotes the GRB duration
classification (short: green; long: pink), while marker edge thickness
scales with the localization association between the GW and GRB sky
positions, with thicker edges indicating stronger spatial consistency.
The sample is temporally asymmetric, with 26 candidates in the
GW-precedes-GRB region and 13 in the GRB-precedes-GW region. No candidate
lies within 1000~s of $\Delta t = 0$, leaving a pronounced gap around
exact temporal coincidence. The mean localization association is also
higher on the GW-precedes-GRB side (36\%) than on the GRB-precedes-GW
side (23\%).
\begin{figure}[!t]
    \centering
    \includegraphics[width=\columnwidth]{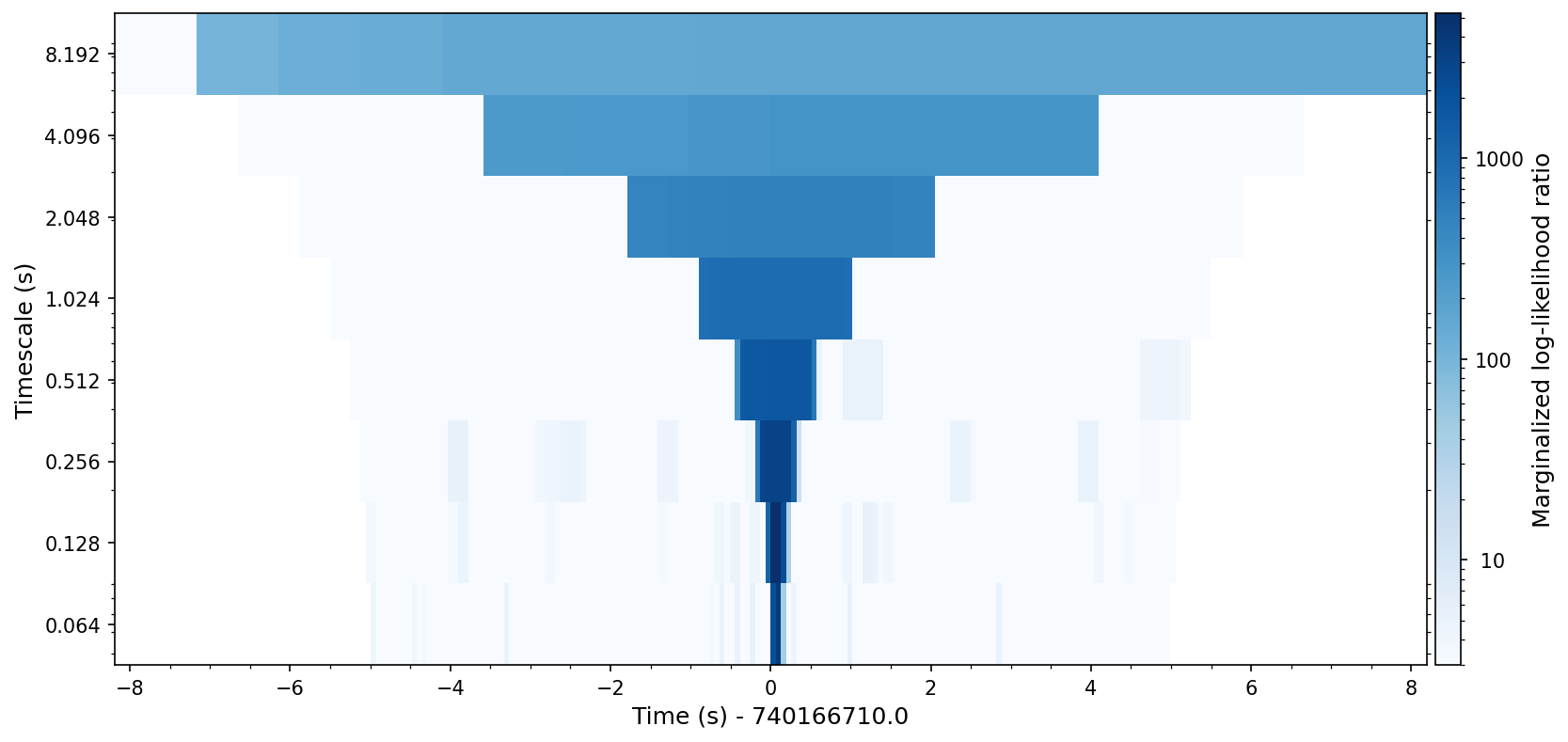}
    \caption{
    Time–timescale map of the marginalized log-likelihood ratio from the
targeted Fermi-GBM search for this GRB candidate. Color intensity
tracks the ratio's strength, with darker blue denoting higher values
and lighter shades denoting lower ones. Two events emerge for this
candidate, with $\log \mathcal{L}_{\rm ratio}=5250.84$ and $23.24$,
respectively. The horizontal axis gives time relative to the GW
trigger, while the vertical axis gives the search duration. Event
properties are listed in Table~\ref{tab:grb_candidates_39}.
    }
    \label{fig:highestloglrwaterfallplot}
\end{figure}
\begin{figure}[!t]
    \centering
    \includegraphics[width=\columnwidth]{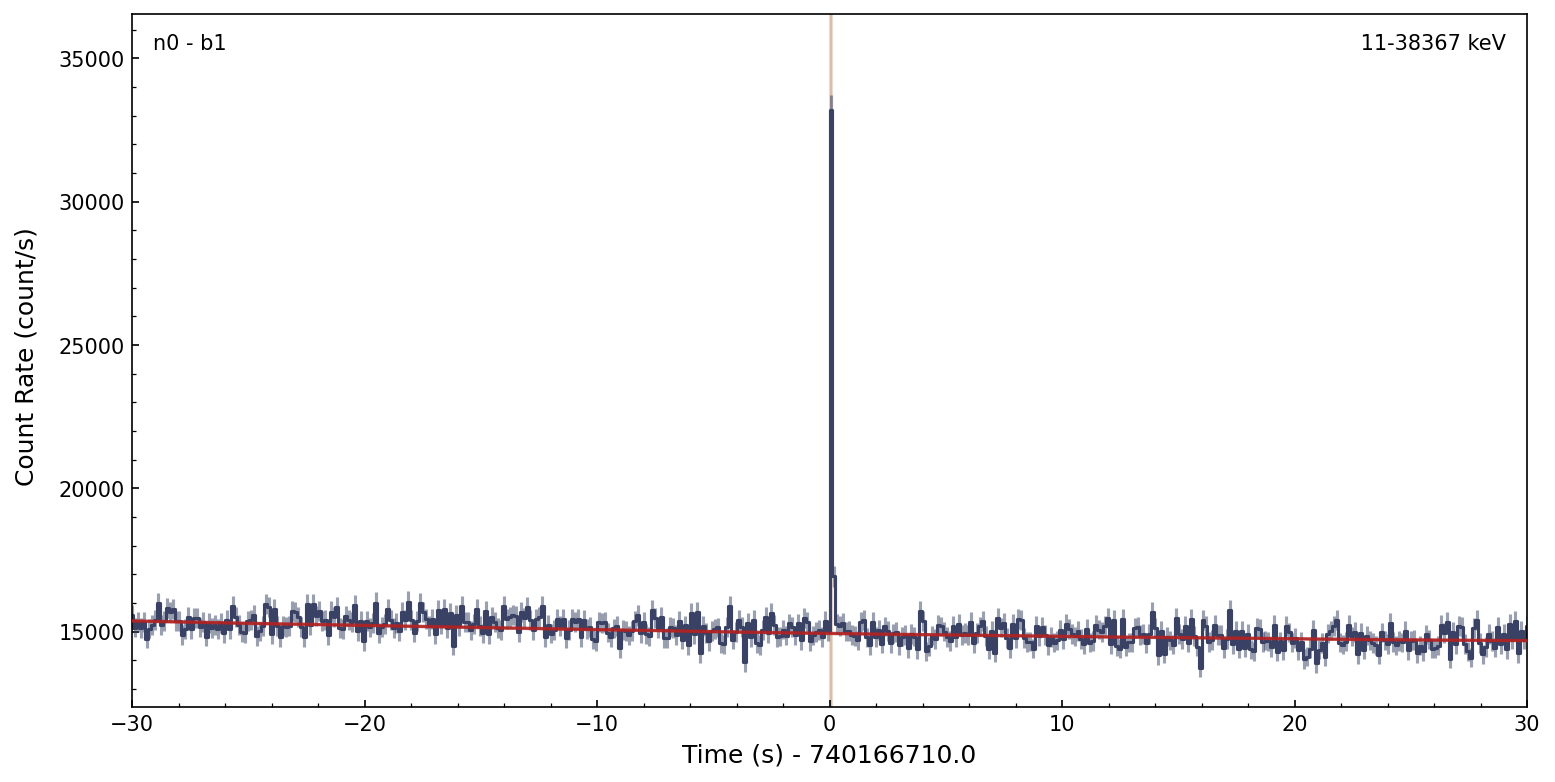}
    \caption{Combined lightcurve summed across all 14 GBM detectors, showing
    the count rate over time around the trigger. The red line shows
    the expected background level, and the sharp spike near $t=0$~s
    marks the candidate signal.
    }
    \label{fig:highestloglrlcplot}
\end{figure}
Figure~\ref{fig:figure3_component_masses} presents the inferred source-frame component masses ($M_1, M_2$) alongside their 90\% credible intervals. Component masses span from $\sim 7\,M_{\odot}$ to $100\,M_{\odot}$, with most events clustering between $10\text{--}50\,M_{\odot}$. Crucially, every component mass lies well above the Tolman--Oppenheimer--Volkoff (TOV) maximum non-rotating neutron star mass ($M_{\text{TOV}} \sim 2\text{--}3\,M_{\odot}$) \citep{Tolman_1939, Oppenheimer_1939, Kalogera_1996, Margalit_2017}, indicating that our sample consists entirely of binary black hole (BBH) mergers with no neutron star components \citep{LVK_2026_GWTC5}. By convention, $M_1 \ge M_2$ across all events, and parameter uncertainties widen at higher total masses due to the shorter in-band signal duration and reduced parameter-estimation precision typical of high-mass systems \citep{Cutler_1994, Veitch_2015}.
Table~\ref{tab:grb_info} summarizes the properties of the deduplicated sub-threshold GRB candidates identified within the 24-hour analysis window of each gravitational-wave event. For each candidate, We report the sub-threshold GRB trigger time ($T_0$), log-likelihood ratio ($\log\mathcal{L}_{\mathrm{ratio}}$), and candidate duration ($T_{\mathrm{cand}}$)---the dyadic search timescale ($0.064$--$8.192\,\mathrm{s}$) at which each candidate is recovered. Candidates are classified as short ($T_{\mathrm{cand}} < 2\,\mathrm{s}$) or long ($T_{\mathrm{cand}} \ge 2\,\mathrm{s}$) based on this recovered duration rather than $T_{90}$. Significances range from the pre-filtering threshold ($\log\mathcal{L}_{\mathrm{ratio}} \sim 5$) up to $\sim 5250$, and the sample is dominated by short-duration candidates.

Table~\ref{tab:gw_grb_assoc} lists the GW--GRB candidate associations within the 24-hour window, with GW events arranged in reverse chronological order (from newest to oldest). For each pairing, we report the localization overlap, temporal delay ($\Delta t$), preceding event status, and the GW component masses ($m_1$, $m_2$). The ID column corresponds directly to the candidate IDs listed in Table~\ref{tab:grb_info}, allowing the detailed properties of each associated GRB candidate to be readily cross-referenced.

\section{Discussion}
\label{sec:discussion}
Across the 39 identified sub-threshold candidates, none meet the physical and observational requirements for a true multi-messenger association. While our targeted search identified several high-ranked sub-threshold candidates, a comprehensive evaluation of their statistical, spatial, temporal, and physical properties. Below, we discuss each line of evidence in detail.

\subsection{Highest Statistical Significance}
\label{subsec:trials_factors}

The targeted search recovered a significant candidate at Fermi-GBM
trigger time $740166710$, with $\log\mathcal{L}_{\text{ratio}} = 5250.84$
and a duration of $0.128$~s, localized to RA $= 328.82^\circ$, Dec $= 47.20^\circ$
The waterfall plot and summed lightcurve for this trigger are shown in
Figure~\ref{fig:highestloglrwaterfallplot} and
Figure~\ref{fig:highestloglrlcplot}, respectively.

An independent short GRB, GRB 240615A, was detected onboard by Fermi-GBM
approximately 5~s later within the same search window, with a reported
flux consistent with a bright, prompt event (GCN Circular 36671). To
assess whether these two events share a common origin, we re-ran the
targeted search using GRB 240615A's own GBM trigger time as $T_0$. The
search recovered this burst with extremely high significance
($\log\mathcal{L}_{\text{ratio}} = 6037.78$, SNR $= 105.58$) and
RA $= 335.29^\circ$, Dec $= 44.68^\circ$ --- within $2.18^\circ$ of the
independent onboard GBM localization (RA $= 335.4^\circ$, Dec $= 42.5^\circ$; GCN
36671), consistent with GBM's typical systematic localization
uncertainty. This validates the localization accuracy of the targeted
search and confirms it correctly and sensitively recovers bright,
well-localized bursts when properly centered in time.

Comparing the two localizations derived from the targeted search
directly, we find an angular separation of $5.15^\circ$ between our
candidate and GRB 240615A --- consistent with the $6.62^\circ$ separation
obtained using GRB 240615A's independently reported GCN position, and
comparable to Fermi-GBM's typical systematic localization uncertainty.
While this separation does not permit definitive spatial exclusion at
high confidence, it is combined here with two independent lines of
evidence: (i) the targeted search demonstrably and sensitively recovers
GRB 240615A when correctly centered on its own trigger time, ruling out
a sensitivity-based non-detection at the offset time, and (ii) the two
events are temporally offset by approximately $5$~s, longer than
typical short-GRB prompt-emission timescales. Taken together, these
results support the interpretation that this candidate represents a
genuine, independent detection, distinct from GRB 240615A.

A remaining open question is why a signal of such high significance
($\log\mathcal{L}_{\text{ratio}} = 5250.84$) and short duration
($0.128$~s) was not identified by Fermi-GBM's onboard triggering
algorithm at trigger $740166710$, warranting further investigation
into the onboard trigger's sensitivity for events of this nature.

\subsection{Spatial Overlap vs. Temporal Offset Constraints}
\label{subsec:spatial_temporal_constraints}
We now consider a separate candidate, GW240930\_035959 (Fermi-GBM
trigger 749409554), which exhibits the highest spatial association
in our sample (Table~\ref{tab:gw_grb_assoc}). The joint localization
association of 95.4\% is higher than the 93.7\% found for the
confirmed event GW170817/GRB~170817A using the same search. This is not simply a result of two large, poorly
constrained regions overlapping by chance: the two localizations
share a well-defined region of $158.3$~deg$^2$
(Figure~\ref{fig:highestassociation}), covering $42.9\%$ of the GW area
and $23.7\%$ of the GBM area. By this measure, the spatial agreement
for GW240930\_035959 is stronger than for the confirmed event.

However, the candidate is offset from the GW merger time by
$\Delta t = 13$~h $19$~min $8.85$~s, far beyond any timescale expected
for prompt emission from a compact binary merger\citep{Yang_2020}. For this reason
alone, we do not classify it as a confirmed multi-messenger
counterpart.
This contrast raises a genuine open question: how should we interpret
a spatial association stronger than a confirmed event, when the same
candidate is ruled out by a single criterion -- timing -- while every
other measure points toward consistency? Two different explanations
are possible.

The first is that the two signals are simply unrelated events whose
localizations happen to overlap by chance. For example, two separate
compact binary mergers, or a compact binary merger and an unrelated
GRB, could each be well localized on their own and still end up
pointing to nearby, or even overlapping, regions of the sky without
any physical connection between them.

The second possibility is that the two events do share a common
origin, but the delay reflects a process we do not yet expect for
compact binary mergers -- for instance, some form of delayed emission
from the merger remnant, occurring long after the initial burst rather
than promptly. This would require an emission mechanism operating on
timescales far longer than the prompt short-GRB window, which is not
established in current models.

We do not attempt to determine which explanation is correct here, as
this analysis considers only spatial and temporal coincidence.
Instead, we present this candidate as motivation for further
investigation, including archival searches for any independent
electromagnetic counterpart to either event.
These two possible explanations also point to two open questions for
future work. First, if unrelated events can produce spatial overlap
this strong purely by chance, how often should we expect this to
happen, and at what point does spatial overlap alone become
meaningful without also checking the timing? Answering this would
require estimating how often two independent, well-localized events
would overlap this much by coincidence. Second, if a genuine but
delayed connection between such events is physically possible, what
kind of emission mechanism could produce a delay on the order of
hours rather than seconds, and could such a signature be
distinguished from an unrelated coincidence using additional
information, such as an independent distance estimate or spectral
comparison? We leave both questions open for future investigation.
\subsection{Progenitor Masses and the TOV Limit}
\label{subsec:mass_tov_constraints}

The inferred source-frame component masses ($m_1, m_2$) of the gravitational wave (GW) triggers in our sample span from $\sim 7\,M_{\odot}$ to $\sim 100\,M_{\odot}$ ($90\%$ credible intervals) \citep{Connaughton_2016, LVK_2026_GWTC5}. Because every component mass significantly exceeds the Tolman--Oppenheimer--Volkoff (TOV) limit for a cold, non-rotating neutron star ($M_{\text{TOV}} \approx 2.2\text{--}2.3\,M_{\odot}$) \citep{Tolman_1939, Oppenheimer_1939, Kalogera_1996, Margalit_2017}, the sample consists entirely of binary black hole (BBH) coalescences.

Standard models dictate that prompt short gamma-ray bursts (GRBs) require a neutron star component whose tidal disruption produces a central accretion disk \citep{Eichler_1989, Narayan_1992, Metzger_2012}. Vacuum BBH mergers lack tidal ejecta and are thus not expected to launch prompt relativistic jets \citep{Connaughton_2016, LVK_2026_GWTC5}. Consequently, these high component masses ($m_1, m_2 \gg M_{\text{TOV}}$) suggest that any coincident sub-threshold gamma-ray burst is a chance alignment rather than a physically associated transient \citep{Kocevski_2018, Hamburg_2020, Cai_2025}.
\subsection{Summary of Candidate Nature}
\label{subsec:discussion_summary}

Taken together, the likelihood-ratio rankings, candidate timescale distributions, temporal delays ($\Delta t > 17\text{ min}$), spatial map properties, and component masses ($M_1, M_2 \gg M_{\text{TOV}}$) lead to a unified conclusion: the recovered candidates are consistent with background-dominated statistical fluctuations occurring by chance within the 24-hour GW search windows, rather than genuine electromagnetic counterparts to compact binary mergers.

\begin{table*}
\centering
\caption{Complete sample of sub-threshold GRB candidates identified in this analysis, provided here as supplementary reference material. For each candidate, we list the trigger identifier, start time ($t_{\rm start}$) relative to the search window, duration, azimuth (az) and zenith (zen) angles, detection status, signal-to-noise ratio (SNR), spectral template index (Spec), amplitude (Ampl), and the log-likelihood ratio ($\log\mathcal{L}_{\rm ratio}$) used to rank candidate significance. This table summarizes the full underlying dataset from which the results discussed in the main text (Section~\ref{sec:results}) are derived.}
\label{tab:grb_candidates_39}
\vspace{0.5em}
{\tiny
\renewcommand{\arraystretch}{1.1}
\begin{tabular*}{\textwidth}{@{\extracolsep{\fill}} c c c c c c c c c c}
\hline\hline
\textbf{Trigger} & \textbf{$t_{\text{start}}$} & \textbf{Duration} & \textbf{az} & \textbf{zen} & \textbf{Status} & \textbf{SNR} & \textbf{Spec} & \textbf{Ampl} & \textbf{$\log\mathcal{L}_{\rm ratio}$} \\
\hline\hline
740166710 & 0 & 0.128 & 200 & 15 & 1 & 100.1 & 0 & 14.09 & 5250.84 \\
 & 0.128 & 0.064 & 240 & 30 & 1 & 7.17 & 1 & 1.93 & 23.24 \\
\hline
759019271 & 0.192 & 0.256 & 160.62 & 65 & 1 & 23.88 & 0 & 3.88 & 284.24 \\
 & -3.392 & 0.064 & 172.17 & 140 & 1 & 6.5 & 1 & 3.88 & 16.69 \\
 & 2.304 & 0.128 & 0 & 20 & 1 & 4.68 & 2 & 0.59 & 5.73 \\
\hline
754776177 & 0.64 & 0.064 & 27.69 & 65 & 1 & 22.44 & 2 & 3.37 & 258.03 \\
\hline
758917876 & 0.896 & 0.128 & 26.18 & 50 & 1 & 22.14 & 0 & 2.9 & 244.18 \\
 & 0.256 & 1.024 & 27.69 & 65 & 1 & 9.12 & 1 & 0.6 & 35.76 \\
\hline
747661355 & 0.384 & 0.256 & 59.1 & 70 & 1 & 15.62 & 1 & 2.27 & 119.94 \\
 & -1.024 & 4.096 & 337.85 & 65 & 1 & 9.02 & 2 & 0.18 & 33.56 \\
 & 1.28 & 0.064 & 322.39 & 110 & 1 & 5.28 & 0 & 1.2 & 10.21 \\
\hline
755576606 & 0.64 & 0.512 & 340.36 & 50 & 1 & 24.66 & 1 & 2.21 & 302.11 \\
 & 0.768 & 0.064 & 352.17 & 40 & 1 & 10.5 & 0 & 1.91 & 51.13 \\
 & 1.088 & 0.512 & 60 & 15 & 1 & 5 & 2 & 0.29 & 8.95 \\
\hline
754570926 & 0.128 & 0.064 & 30.42 & 95 & 1 & 12.68 & 2 & 2.33 & 78.87 \\
 & 2.496 & 0.064 & 220 & 15 & 1 & 3.7 & 0 & 0.72 & 5.43 \\
\hline
747000169 & 0.576 & 0.256 & 255 & 20 & 1 & 12.8 & 1 & 1.43 & 77.65 \\
 & 0.384 & 0.256 & 111.72 & 55 & 1 & 7.86 & 2 & 0.64 & 26.21 \\
 & 1.024 & 0.064 & 184.39 & 35 & 1 & 3.67 & 0 & 0.54 & 5.09 \\
\hline
746731612 & 0.768 & 0.128 & 210.86 & 80 & 1 & 11.95 & 1 & 2.25 & 63.94 \\
 & -1.216 & 0.128 & 0 & 100 & 1 & 4.33 & 2 & 0.49 & 5.26 \\
\hline
749409554 & -6.144 & 8.192 & 240 & 105 & 1 & 12.13 & 2 & 0.22 & 63.3 \\
 & 0.256 & 0.512 & 252 & 25 & 1 & 5.61 & 2 & 0.41 & 10.99 \\
\hline
746684112 & 0.256 & 0.128 & 10.43 & 105 & 1 & 10.04 & 0 & 1.62 & 46.29 \\
 & -4.736 & 0.064 & 299.15 & 95 & 1 & 4.9 & 2 & 0.79 & 7.52 \\
 & 4.032 & 0.512 & 91.64 & 50 & 1 & 4.54 & 1 & 0.38 & 5.85 \\
\hline
746197881 & -1.536 & 4.096 & 145 & 90 & 1 & 10.22 & 1 & 0.32 & 44.81 \\
 & 1.536 & 4.096 & 172.39 & 85 & 1 & 5.65 & 2 & 0.11 & 10.5 \\
 & 1.408 & 0.128 & 260 & 150 & 1 & 4.37 & 0 & 1.4 & 5.75 \\
\hline
748190210 & 0.064 & 0.064 & 348 & 25 & 1 & 9.19 & 0 & 1.74 & 41.46 \\
 & 0 & 0.064 & 116.31 & 65 & 1 & 4.03 & 2 & 0.75 & 5.2 \\
\hline
748985632 & 0.96 & 0.064 & 98.71 & 120 & 1 & 8.46 & 2 & 1.94 & 35.84 \\
 & 2.56 & 4.096 & 272.9 & 60 & 1 & 4.09 & 0 & 0.08 & 5.14 \\
\hline
757181763 & -1.024 & 8.192 & 288 & 155 & 1 & 7.55 & 0 & 0.34 & 22.66 \\
 & -0.256 & 0.256 & 313.04 & 75 & 1 & 5.29 & 2 & 0.39 & 10.15 \\
\hline
741848134 & -0.512 & 2.048 & 60 & 90 & 1 & 7.02 & 2 & 0.21 & 18.88 \\
 & 0.448 & 0.128 & 7.83 & 140 & 1 & 4.14 & 0 & 0.68 & 7.25 \\
 & -2.048 & 2.048 & 58.06 & 120 & 1 & 4.65 & 1 & 0.3 & 5.88 \\
\hline
754439562 & 0.768 & 0.128 & 67.83 & 105 & 1 & 7.18 & 1 & 1.48 & 18.4 \\
 & -0.512 & 0.128 & 96.72 & 70 & 1 & 3.98 & 0 & 0.46 & 5.28 \\
\hline
746739456 & 0.64 & 0.064 & 126.76 & 85 & 1 & 5.51 & 1 & 1.46 & 12.94 \\
\hline
748414684 & -0.128 & 0.512 & 180 & 60 & 1 & 5.65 & 1 & 0.48 & 12.03 \\
\hline
748828348 & -0.512 & 2.048 & 31.3 & 75 & 1 & 5.27 & 0 & 0.14 & 10.46 \\
\hline
758135659 & 0.704 & 0.064 & 180 & 10 & 1 & 5.37 & 1 & 1.46 & 10.39 \\
 & -4.8 & 0.064 & 0 & 120 & 1 & 4.6 & 0 & 0.82 & 6.02 \\
\hline
754251071 & -1.024 & 2.048 & 204.83 & 125 & 1 & 5.3 & 1 & 0.34 & 9.01 \\
 & -1.856 & 0.064 & 279.4 & 70 & 1 & 4.34 & 2 & 0.63 & 6.97 \\
\hline
741908097 & -0.512 & 4.096 & 30.86 & 80 & 1 & 5.24 & 1 & 0.17 & 8.78 \\
\hline
742852188 & 0.576 & 0.064 & 28.8 & 45 & 1 & 4.56 & 0 & 1.26 & 7.96 \\
\hline
734754392 & -0.384 & 0.064 & 235.86 & 55 & 1 & 5.28 & 2 & 0.66 & 7.95 \\
 & -1.28 & 2.048 & 0 & 0 & 1 & 4.98 & 1 & 0.24 & 7.52 \\
\hline
746311162 & 0.832 & 0.064 & 126.76 & 85 & 1 & 4.85 & 1 & 1.92 & 7.49 \\
 & 0.32 & 0.256 & 0 & 35 & 1 & 4.5 & 0 & 0.31 & 6.62 \\
\hline
738941649 & 0.576 & 0.064 & 148.7 & 40 & 1 & 4.24 & 1 & 1.49 & 7.27 \\
 & 4.864 & 0.128 & 50.7 & 85 & 1 & 4.68 & 0 & 0.66 & 7.19 \\
 & 2.56 & 4.096 & 213.91 & 105 & 1 & 4.68 & 2 & 0.12 & 6.59 \\
\hline
758771314 & 0.384 & 0.064 & 262.29 & 80 & 1 & 4.04 & 0 & 0.89 & 7.18 \\
\hline
748992037 & 0.768 & 0.128 & 188.31 & 115 & 1 & 3.93 & 0 & 0.61 & 6.63 \\
\hline
748640542 & 0.512 & 0.512 & 300 & 25 & 1 & 4.76 & 1 & 0.45 & 6.1 \\
 & 1.28 & 0.128 & 0 & 35 & 1 & 4.26 & 0 & 0.56 & 5.06 \\
\hline
741891537 & -1.024 & 4.096 & 109.57 & 75 & 1 & 4.71 & 1 & 0.16 & 5.81 \\
\hline\hline
\end{tabular*}
}
\end{table*}

\section{Acknowledgement}
The author gratefully acknowledges the public data and software provided by the \textit{Fermi} Gamma-ray Burst Monitor (Fermi-GBM) team and the NASA High Energy Astrophysics Science Archive Research Center (HEASARC). This work also makes use of public gravitational-wave candidate catalogs and data provided by the LIGO Scientific Collaboration, the Virgo Collaboration, and the KAGRA Collaboration (LVK). 

The author extends sincere thanks to Dr.~Amit Shukla, Dr.~Sayantan Bhattacharya, and Chandan Kumar Das for valuable discussions, guidance, and constructive feedback throughout the development of this work.

\bibliographystyle{aasjournalv7}
\bibliography{sample701}
\clearpage

\clearpage

\end{document}